# Efficient coupling of within- and between-host infectious disease dynamics

Cameron A. Smith[1,4,*] and Ben Ashby[2,3,4]

1 Department of Biology, University of Oxford, Oxford, UK

2. Department of Mathematics, Simon Fraser University, Burnaby, BC, Canada

3. Pacific Institute on Pathogens, Pandemics and Society, Burnaby, BC, Canada

4. Department of Mathematical Sciences, University of Bath, Bath, UK

*Corresponding author: cameron.smith@biology.ox.ac.uk

**Keywords:** epidemiology, compartmental models, multiscale, hybrid models, nested models

## Abstract

Mathematical models of infectious disease transmission typically neglect within-host dynamics. Yet within-host dynamics – including pathogen replication, host immune responses, and interactions with microbiota – are crucial not only for determining the progression of disease at the individual level, but also for driving within-host evolution and onwards transmission, and therefore shape dynamics at the population level. Various approaches have been proposed to model both within- and between-host dynamics, but these typically require considerable simplifying assumptions to couple processes at contrasting scales (e.g., the within-host dynamics quickly reach a steady state) or are computationally intensive. Here we propose a novel, readily adaptable and broadly applicable method for modelling both within- and between-host processes which can fully couple dynamics across scales and is both realistic and computationally efficient. By individually tracking the deterministic within-host dynamics of infected individuals, and stochastically coupling these to continuous host state variables at the population-level, we take advantage of fast numerical methods at both scales while still capturing individual transient within-host dynamics and stochasticity in transmission between hosts. Our approach closely agrees with full stochastic individual-based simulations and is especially useful when the within-host dynamics do not rapidly reach a steady state or over longer timescales to track pathogen evolution. By applying our method to different pathogen growth scenarios we show how common simplifying assumptions fundamentally change epidemiological and evolutionary dynamics.

## Significance statement

Within-host dynamics – often neglected in traditional epidemiological and evolutionary modelling – are important for determining the individual and population level outcomes of infectious diseases. Here we introduce a multiscale method for modelling infectious diseases at between- and within-host levels which fully captures the dynamics at both scales. This method has significant consequences for modelling the epidemiology and evolution of infectious diseases as it allows us to efficiently investigate the effects of within-host processes on disease transmission and contrasting selection pressures at different scales, as well as to simulate exact phylogenetic trees.

# INTRODUCTION

Since their inception nearly a century ago, compartmental models of infectious diseases have proved to be incredibly powerful for modelling epidemiological dynamics while requiring relatively few parameters and assumptions [1,2]. Compartmental models are not only widely used for understanding and predicting disease transmission, but also provide crucial insights into fundamental evolutionary processes in pathogens, including selection for virulence [3], immune evasion [4], and anti-microbial resistance [5]. Such models also shed light on pathogen coevolution with their host populations, including humans [6], animals [7], plants [8], fungi [9], and bacteria [10]. The development of realistic and efficient modelling frameworks has therefore been identified as a key goal for the modelling community, especially those that link processes across scales [11].

For simplicity, compartmental models typically only focus on epidemiological dynamics at the between-host level (e.g., transmission) and neglect within-host processes (e.g., pathogen growth and immune response dynamics). This greatly simplifies the difficulty of modelling nonlinear dynamics in large populations as one can classify hosts into distinct groups according to their current disease status (e.g., "susceptible", "infected", or "recovered") without needing to track individual pathogen dynamics within each infected host. Movement between disease-status compartments occurs due to processes such as infection and recovery that do not depend on individual within-host properties such as pathogen load or T-cell count. When fitting to real-world data, this means that key disease parameters, such as the basic reproduction number or serial interval, may be inferred from population-level data without knowledge of the dynamics that occur within hosts.

Although between-host compartmental models give vital insights into pathogen epidemiology and evolution, within-host dynamics including pathogen replication and growth, the host immune response, pharmacokinetics and pharmacodynamics, and interactions with the microbiome and other pathogens are also known to play a critical role [12]. For example, transmission and virulence are both likely to depend on pathogen load. Moreover, theoretical studies have shown that within-host dynamics can play a crucial role in pathogen evolution, leading to immune escape [13] or more varied predictions for virulence evolution [14]. Dynamics at the within-host level can clearly have important implications at the population-level, which can be captured using nested or hybrid models that couple dynamics at the intra- and inter-host scales [12].

While multiple approaches to nested models have been developed [12], these methods are not widely used because tracking the within-host dynamics of all infected individuals in a population is computationally intensive and analytically intractable. For example, in a standard *SIR* model lacking within-host dynamics, only three equations are required to track the population dynamics, whereas a full nested model requires at least the same number of equations as infected individuals. Moreover, if the transmission rate depends on pathogen load, which varies through time at the individual level, then each infected host has a unique time-dependent transmission rate. Many nested models therefore assume that within-host processes rapidly reach a steady state as this eliminates the need to track transient within-host dynamics. However, this separation of timescales does not capture the impact of transient within-host dynamics on transmission and virulence and fails when the within-host dynamics are slower or do not reach a steady state (e.g., acute infections). Stochastic individual-based models (IBMs) [15–18] or integro-differential equations (IDEs) [19–22] have been used to fully capture transient dynamics, but IBMs are computationally intensive, and IDEs can only be used when the within-host dynamics are relatively simple. Methods that fully capture transient within-host dynamics and are computationally efficient are currently lacking.

Here, we propose a novel hybrid method for coupling transient within- and between-host dynamics that is both computationally efficient and realistic. Our method leverages different modelling paradigms across scales to track individual-level within-host dynamics for a discrete number of infected hosts, which are then stochastically coupled to continuous population-level processes. We illustrate our method with contrasting examples of within-host pathogen dynamics, showing the critical impact on epidemiological and evolutionary dynamics at the population level.

## METHODS

We introduce a novel method for coupling infectious disease dynamics within individual hosts with population-level processes. For simplicity, we focus on a single directly transmitted pathogen in a homogeneous, randomly mixing host population, but our approach may be readily generalised for multiple co-circulating pathogens or symbionts, alternative modes of transmission (e.g., vector-borne or sexually transmitted), or for heterogeneous or structured populations. The population consists of $N(t)$ individuals at time $t$. Each host has a given disease state (e.g., susceptible, infected, or recovered), with the density of each state tracked at the population level (*see Host demographic dynamics*; HDD), and a within-host state (e.g., pathogen load, T-cell count), where applicable (see *Within-host dynamics*; WHD*)*. We use systems of ordinary differential equations (ODEs) to model the dynamics at each scale for their simplicity and broad applicability, but our framework could be easily applied to other approaches, including difference equations, stochastic differential equations, or stochastic simulation algorithms. To couple the different scales efficiently, we convert between a discrete number of hosts at the WHD level and a continuous density of hosts at the population-level (see *Coupling* and schematic in Fig. 1). Below, we first describe general frameworks for modelling the within-host dynamics of an infectious disease and the demographic dynamics of the host population, before introducing our novel hybrid method for coupling dynamics across these two scales.

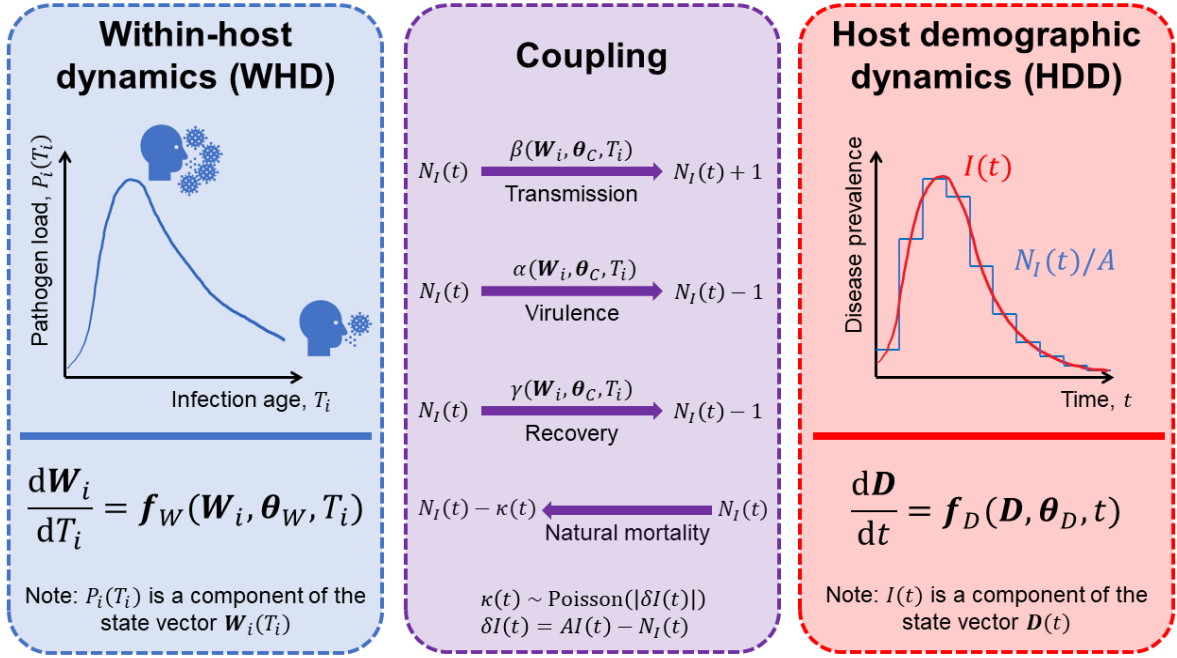

*Fig. 1: **Model schematic for the hybrid method.** The within-host dynamics (WHD) track pathogen replication and immune dynamics for each infected host, i at infection time $T_i$, in state vector $W_i(T_i)$ for a discrete number of infected hosts. The WHD have instantaneous rate of change $f_W(W_i, \theta_W, T_i)$, where $\theta_W$ is a vector of within-host parameters. The host-demographic dynamics (HDD) track continuous densities of population-level disease states and include births, natural mortality and migration in state vector $D(t)$, which changes according to instantaneous rate of change $f_D(D, \theta_D, t)$, where $\theta_D$ is a vector of host demographic parameters and t is the population-level time. The coupling functions link these two scales together stochastically through epidemiological processes (transmission, mortality virulence, and recovery) to update the host disease states at the HDD level, and by removing infected individuals at the WHD level due to natural mortality.*

*WITHIN-HOST DYNAMICS*

The within-host dynamics (WHD) of an infectious disease may include replication and growth of the pathogen population from an initial infectious dose, the host immune response and its interactions with the pathogen, positive or negative interactions with other pathogens or constituents of the host microbiome, and within-host evolution of the pathogen through selection or drift. To model the WHD, we assign a unique set of ordinary differential equations (ODEs) to every individual in the population (although in practice one might limit the within-host dynamics only to currently infected or recently recovered individuals).

Consider an individual ($i$) who was infected $T_i$ time-units ago. At time $T_i$, their within-host state is given by the function $\boldsymbol{W_i}(\boldsymbol{\theta}_W, T_i)$, where $\boldsymbol{\theta}_W$ contains all parameters that are required for the evaluation of the within-host state. These parameters could include, but are not limited to, rates for pathogen replication and the upregulation of an immune response. The within-host dynamics are then given by:

$$\frac{d\boldsymbol{W}_i}{dT_i} = \boldsymbol{f}_W(\boldsymbol{W}_i, \boldsymbol{\theta}_W, T_i), \quad \boldsymbol{W}_i(\boldsymbol{\theta}_W, 0) = \widetilde{\boldsymbol{W}}(\boldsymbol{\theta}_W). \tag{1}$$

where $\boldsymbol{f}_W(\boldsymbol{W}_i, \boldsymbol{\theta}_W, T_i)$ describes how the within-host states change subject to the initial condition $\widetilde{\boldsymbol{W}}(\boldsymbol{\theta}_W)$, which could be a function of the infecting individual (see *Coupling*).

*HOST DEMOGRAPHIC DYNAMICS*

At the population level, we distinguish host demographic dynamics (HDD), such as births and background mortality, from disease-related processes such as transmission, recovery and disease-associated mortality (virulence), which are captured through the coupling of the two scales (see

*Coupling*). In principle, one could include other processes such as migration or mating dynamics (for non-sexually transmitted pathogens in animal populations) at this scale.

As with the WHD, one can model the HDD stochastically, or deterministically using the following general ODEs:

$$\frac{d\boldsymbol{D}}{dt} = \boldsymbol{f}_D(\boldsymbol{D}, \boldsymbol{\theta}_D, t), \quad \boldsymbol{D}(\boldsymbol{\theta}_D, 0) = \widetilde{\boldsymbol{D}}(\boldsymbol{\theta}_D), \tag{2}$$

Where the host demographic state at time $t$ is given by $\boldsymbol{D}(\boldsymbol{\theta}_D, t)$ with parameter set $\boldsymbol{\theta}_D$ and initial condition $\widetilde{\boldsymbol{D}}(\boldsymbol{\theta}_D)$, and the function $\boldsymbol{f}_D(\boldsymbol{D}, \boldsymbol{\theta}_D, t)$ describes the change in the density of each host state due to non-disease related processes (e.g., births and background mortality). Parameters at this scale would include, for example, rates for host birth and death, or migration.

## COUPLING

We now describe coupling of the within-host (WHD) and host demographic (HDD) dynamics, together with the change in host state due to disease-related processes (infection, recovery, death) (Fig. 1). Recall that hosts are tracked individually for the WHD, but the HDD use population densities. We focus on infected individuals, but the same approach could be applied to any host state. The density of infected hosts at time $t$, $I(t) \in \mathbb{R}_+$, is given by $I(t) \approx N_I(t)/A$, where $N_I(t) \in \mathbb{N}$ is the number of infected individuals at time $t$ and $A$ is a positive constant representing the area over which individuals interact. Note that continuous compartmental models of infectious diseases track population densities rather than population sizes [1]. We use stochastic updates to determine the number of times each event takes place over a small time-interval $[t, t + \Delta)$. The HDD state variables and the discrete individuals are then updated simultaneously. We focus on infection, recovery, and virulence as disease-related processes, but the method could be readily extended to other processes (e.g., transitions to or from latency, or between asymptomatic and symptomatic disease).

The general transmission function, $\beta(\mathbf{W}, \boldsymbol{\theta}_C, T)$, describes the rate at which a host with infection age $T$ and within-host state $\mathbf{W}$ transmits the pathogen to any individual currently susceptible to the infection, where $\boldsymbol{\theta}_C$ denotes any coupling parameters (e.g., defining the relationship between pathogen load and transmission). We use the function $\beta(\mathbf{W}, \boldsymbol{\theta}_C, T)$ to stochastically calculate the number of new infections arising from each infected individual in a short period of time, $[T, T + \Delta)$, using the $\tau$-leap method [23]. The number of new infections produced by infected individual $i$ during this time interval is drawn from the probability distribution

$$n_i(t) \sim Poisson(\beta(\mathbf{W}_i, \boldsymbol{\theta}_C, T_i) S(t) \Delta). \qquad (3)$$

We then update the number of infected individuals, $N_I(t)$, and the population-level host states, $\mathbf{D}(t)$, accordingly by a mass equivalent to one individual ($1/A$) from the susceptible to the infected state:

$$S(t + \Delta) = S(t) - \frac{1}{A} \sum_{i=1}^{N_I(t)} n_i(t), \qquad (4)$$

$$I(t + \Delta) = I(t) + \frac{1}{A} \sum_{i=1}^{N_I(t)} n_i(t). \qquad (5)$$

The rate at which an individual with infection age $T$ recovers is given by $\gamma(\mathbf{W}, \boldsymbol{\theta}_C, T)$. For example, the recovery rate may depend on the current pathogen load and T-cell count, or for simplicity may be treated as a constant. To determine if an infected individual $i$ has recovered in this time interval, we draw a uniform random number between 0 and 1, $u_{r,i}$, and if $u_{r,i} < \gamma(\mathbf{W}_i, \boldsymbol{\theta}_C, T_i) \Delta$ then the individual has recovered and we update the population-level and within-host states accordingly and no longer track their within-host dynamics. We use a similar process to determine if an individual has died from disease during this time interval, where the disease-associated mortality rate is

$\alpha(\boldsymbol{W}, \boldsymbol{\theta}_C, T)$. We update the number of infected individuals $N_I(t + \Delta)$ accordingly, and the population-level states by subtracting $1/A$ (the density of a single individual) from the infected state, $I(t)$, for each recovery or mortality event, and adding $1/A$ per recovery event to the recovered or susceptible states depending on the assumptions regarding recovery.

The final stage of the coupling reconciles changes in the HDD with the number of individuals tracked for the WHD. For example, infected individuals may die due to other causes, or the number of infected individuals may change due to migration. We therefore need to ensure that the number of individuals at the WHD level is approximately equivalent to the appropriate HDD state variable. For simplicity, we assume that only infected individuals are tracked for the WHD and that the HDD may result in non-disease-associated mortality, but an analogous approach would apply to other scenarios.

Recall that the relationship between the continuous density of infected individuals in the HDD, $I(t)$, and the discrete number of infected individuals in the WHD, $N_I(t)$, is given by $I(t) \approx N_I(t)/A$. If $\delta I(t) = AI(t) - N_I(t) > 0$ then the number of tracked individuals in the WHD is fewer than the number of individuals according to the HDD. Conversely, if $\delta I(t) < 0$, then the number of tracked individuals is greater than in the HDD. This discrepancy is corrected probabilistically. We draw a Poisson random variable $\kappa(t) \sim Poisson(|\delta I(t)|)$ which determines the number of tracked individuals to either add ($\delta I(t) > 0$) or remove ($\delta I(t) < 0$) at random. In our simulations, we cannot have $\delta I(t) > 0$ due to the absence of migration. When $\delta I(t) < 0$, we choose $\kappa(t)$ infected individuals at random to be removed due to non-disease-associated mortality.

# APPLICATIONS

We illustrate possible applications of our novel coupling method for two pathogen growth scenarios. In the first case, we assume that following infection the pathogen load grows logistically to a steady state and is not cleared by an immune response, whereas in the second case we assume that the pathogen load grows exponentially for a short period before decaying to a lower level due to a build-up of immunity within the host. For simplicity of illustration (but not a limitation of the framework), we do not explicitly model the immune response, but the dynamics of T cells, for example, could readily be incorporated into the WHD. The pathogen load dynamics presented below are qualitatively similar to models with explicit immunity. We compare the full transient dynamics for pathogen load with a steady state approximation, which is commonly used in hybrid models of within- and between-host dynamics [12,20,24,25]. We also demonstrate how our modelling framework can be used to study pathogen evolution and construct phylogenetic trees.

For our first example, the within-host pathogen load grows logistically with growth rate $r$ and carrying capacity $K$. This may be a reasonable model for infections that are not cleared by the immune system, such as certain sexually transmitted pathogens such as *Chlamydia trachomatis* and opportunistic pathogens in immunocompromised hosts (e.g., *Pseudomonas aeruginosa* infections in cystic fibrosis patients), and for non-pathogenic gut microbiota. We assume that the HDD feature a density-dependent birth term and that all individuals have the same background mortality rate, regardless of their disease status. We include two epidemiological coupling functions: one for transmission (which we assume to be proportional to pathogen load); and one for virulence (which we assume is proportional to the square of the pathogen load). We also assume that there is no recovery from infection. A full description of the model is given in the *Supplementary material, Section S1*. Unsurprisingly, both the epidemic growth rate and the size of the epidemic is typically much lower when the transient WHD are taken into account compared to the steady state approximation, especially when the within-host growth rate $(r)$ is slow relative to the average infectious period (Fig. 2). This is because the transmission rate increases with pathogen load and the steady state approximation therefore overestimates transmission, whereas the transient WHD accounts for a period of relatively low transmission early in the infection.

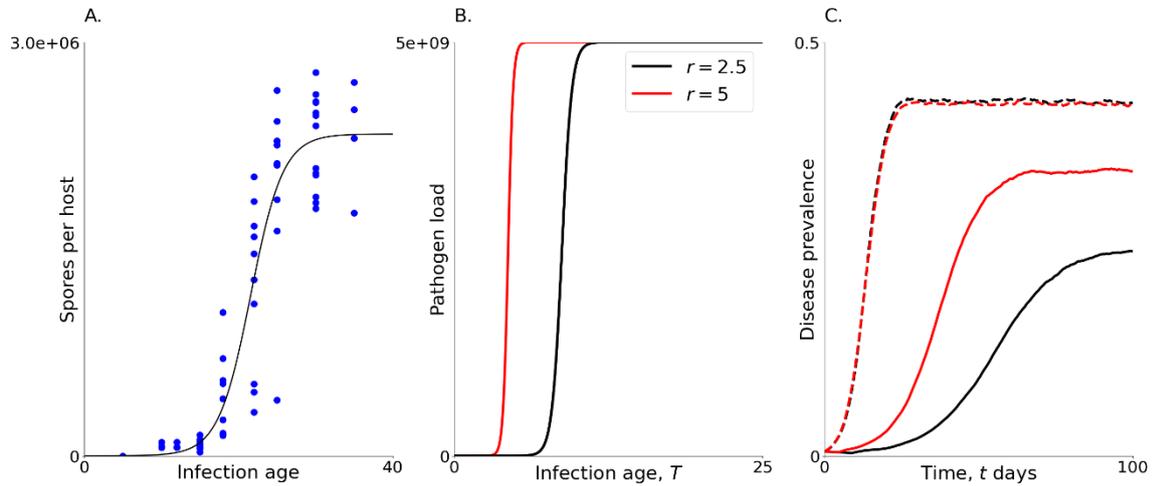

*Fig. 2: **Epidemiological dynamics with logistic pathogen growth.** A. Logistic curve fitted to within-host growth of the bacterial pathogen Pasteuria ramosa in water fleas (Daphnia dentifera) from [26]. B. WHD of pathogen load with relatively slow ($r = 2.5$; black) and fast ($r = 5$; red) pathogen load growth rates. C. Disease prevalence in the host population for relatively slow ($r = 2.5$; black) and fast ($r = 5$; red) pathogen load growth rates, with a steady-state approximation (dashed) or using the full transient dynamics with our novel coupling method (solid). Other parameters as detailed in the Supplementary Material (section S1.5).*

In our second example, the pathogen load initially grows exponentially before peaking and decaying due to an immune response. This is a reasonable approximation for many acute infections, including COVID-19, influenza, and measles, and for the early stages of many chronic infections, such as Syphilis and HIV, where pathogen load peaks soon after infection then falls to lower levels for a long period of time before eventual resurgence. The WHD for pathogen load are controlled by three parameters: $r$, which scales the overall pathogen load; $\eta$, which controls the strength of pathogen decay due to an immune response; and $P^*$, which is the pathogen load steady state. For comparison with the steady state approximation, we assume that the steady state is non-zero. The transmission and virulence functions are the same as in the previous example, but now we also include a recovery rate, which increases exponentially as pathogen load falls. If an individual recovers, they gain full immunity and cannot be reinfected, and their pathogen load is set to zero. A full description of the model can be found in *Supplementary Material* Section S2.

As in the previous example, modelling the full WHD using our coupling method leads to stark differences in epidemic dynamics compared to a steady state approximation (Fig. 3). Since the steady state is relatively low compared to peak pathogen load, this approximation neglects a significantly higher transmission rate during the early stages of infection, and therefore underestimates both the size and growth rate of the epidemic. Naturally, this effect will be stronger the greater the difference between the peak pathogen load and the steady state, and when most transmission occurs during the early stages of an infection.

Not only is our method more realistic than a steady-state approximation, if is also much more computationally efficient than a full stochastic simulation algorithm (SSA). For example, the epidemic simulations in Fig. 2 were approximately 3.5 times faster using our modelling framework than a classical SSA (see *Supplementary Material* Section SX), and 1.7 times faster for the simulations in Fig. 3.

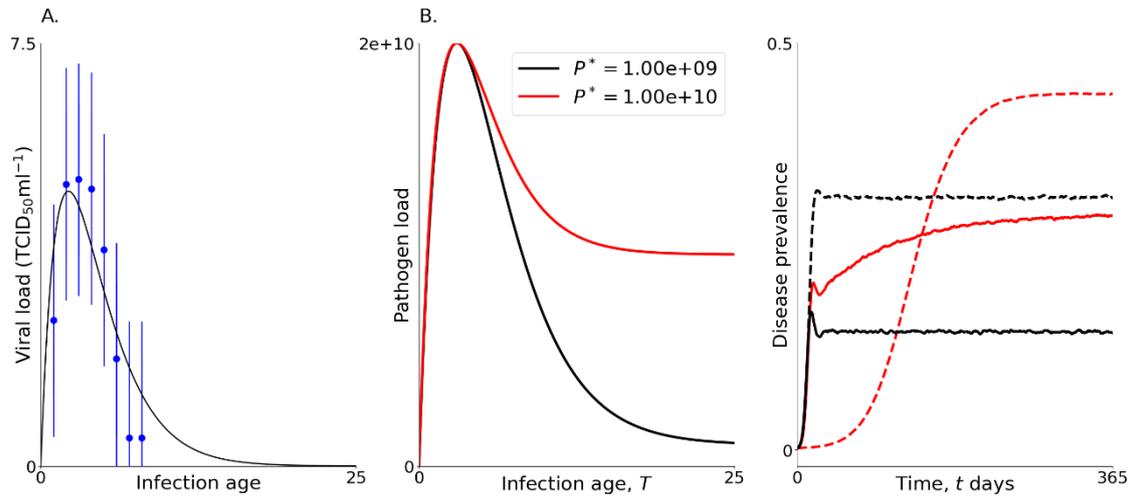

*Fig. 3: **Epidemiological dynamics when pathogen load peaks and decays.** A. Pathogen load (equation (S4); black) fitted to the within-host viral load of averaged influenza A cases (blue) from [27]. B. WHD of pathogen load with relatively low ($P^* = 10^9$; black) and high ($P^* = 10^{10}$; red) long-term pathogen loads. C. Disease prevalence in the host population for relatively low ($P^* = 10^9$; black) and high ($P^* = 10^{10}$; red) long-term pathogen loads at the within-host level, with a steady-state approximation (dashed) or using the full transient dynamics with our novel coupling method (solid). Other parameters as detailed in the Supplementary Material (section S2.5).*

Our modelling framework can also be applied to understand the effects of WHD on pathogen evolution and patterns that emerge at the population level. For example, suppose we consider the evolution of the pathogen growth rate, $r$, in our first example (for a full description of the following simulations, see *Supplementary Material*, Section S1.4). In the steady state approximation, the growth rate has no effect on pathogen fitness, which only depends on the carrying capacity, $K$. The pathogen therefore only experiences drift rather than selection, resulting in a random walk through trait space (Fig. 4A). In contrast, when the transient WHD are modelled using our framework, the pathogen growth rate impacts on transmission and virulence, and therefore fitness, leading to selection towards an evolutionarily stable relationship between pathogen load, transmission, and virulence (Fig. 4B).

As our method tracks infected individuals, along with who infects whom, it can also be readily used to construct true phylogenetic or transmission trees from epidemic simulations with perfect data on infection histories (Fig. 5; *Supplementary Material section S2.4*). This not only has potential uses for efficiently simulating realistic epidemics for testing novel transmission tree reconstruction methods [28], but could also provide new insights into how evolution driven by within-host dynamics impacts the relationship between transmission trees and phylogenetic trees at the population level.

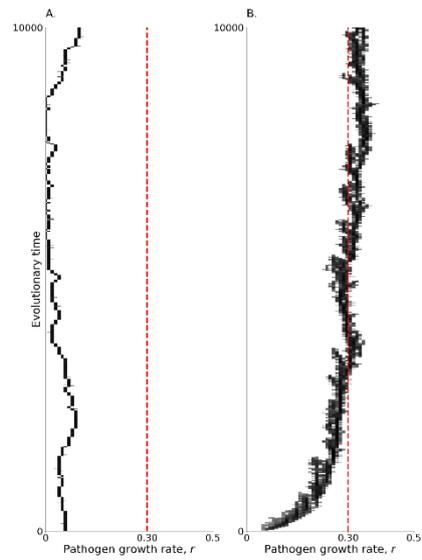

*Fig. 4:* **Evolutionary simulations with logistic pathogen growth.** *A. Evolutionary simulation using the steady-state approximation for the WHD. B. Evolutionary simulation using transient WHD. The red line denotes the evolutionarily stable level of virulence. All simulations conducted using 500 individuals, with density-dependent term $q = 9.9 \times 10^{-4}$. Other parameters as detailed in the Supplementary Material (section S1.5).*

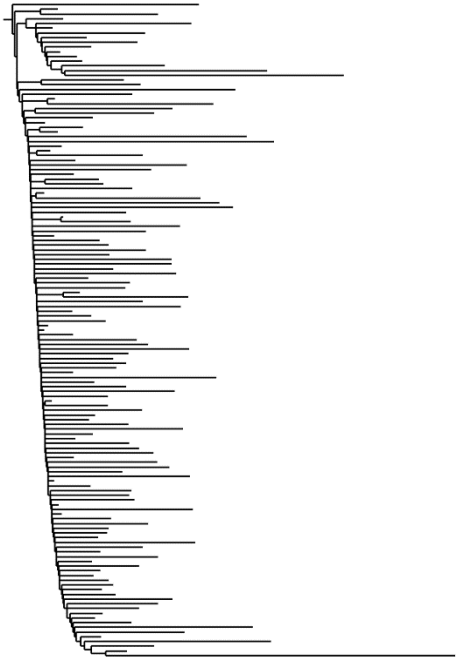

*Fig. 5:* ***Example phylogenetic tree for an evolving pathogen arising from our simulation framework.*** *We have initialised a short length genome (150b in length) and have allowed it to mutate neutrally throughout a population of size 5000. This is the true phylogenetic tree, meaning that all mutation events occurred at the times indicated, rather than a tree reconstruction. Branch length is proportional to time (arbitrary units).*

# DISCUSSION

Understanding how within-host processes impact on the epidemiological and evolutionary dynamics of infectious diseases is a key objective for the infectious disease community. Yet fully capturing transient within-host dynamics in nested models is a major challenge due to the computational burden of tracking individual infection processes, and so many studies assume a separation of timescales so that the within-host dynamics rapidly reach steady state. We have developed a novel hybrid method for efficiently and realistically modelling infectious disease dynamics across scales, capturing both population-level processes (e.g., births and deaths) and individual disease-related processes (e.g., pathogen growth and immune responses). By leveraging fast deterministic models at the population level and for each infected host, and stochastically coupling these scales to enable changes in host states (e.g., susceptible to infected), we make significant speed improvements in even the simplest of models while still fully capturing transient within-host dynamics. We illustrated the power of our framework with two examples of within-host pathogen growth, showing how it can be readily used to model both epidemiological and evolutionary dynamics, often leading to contrasting outcomes compared to widely used steady-state approximations.

A critical feature of our framework is that hosts are tracked as discrete entities for the within-host dynamics, but host states are tracked continuously for the population level dynamics. At the within-host level, each infected individual has their own unique set of ODEs to describe changes in internal states such as pathogen load and T-cell count, but at the population-level host states are tracked continuously using a density-based approximation with a single set of ODEs. This approach allows us to leverage fast ODE-solvers at both scales, which are then coupled using a stochastic tau-leaping method to keep the discrete and continuous levels closely aligned. However, the modular nature of our method means that it can be readily adapted to incorporate different types of models. For example, each of these different scales could be modelled using partial or stochastic differential equations, or with stochastic simulation algorithms. The only requirement is that the coupling correctly aligns the number of individuals tracked for the within-host dynamics with the density of the corresponding host states at the population level. Thus, when an individual is infected, for example, the number of individuals tracked at the within-host level is increased by one, and one unit of host "mass" is moved from the susceptible to infected state at the population-level. Naturally, this means that there will usually be slight discrepancies between the two scales (as illustrated in Fig. 1C), but the density of infected hosts never differs by more than one unit of host mass.

Our method was inspired by hybrid methods for modelling particle diffusion and interaction across a spatial domain with large disparities in particle number. Such spatial hybrid methods [29] employ different modelling paradigms across the spatial domain, with regions containing large numbers of particles using coarser continuum approaches (e.g., PDEs), while low copy number regions use stochastic methods (e.g., agent-based) where continuum limits are invalid. While the two approaches are related in terms of how they handle contrasting scales, our hybrid epidemiological framework uses distinct modelling paradigms for different processes, as opposed to modelling the same phenomena using disparate modelling paradigms.

We demonstrated our method for two contrasting scenarios for within-host pathogen dynamics – logistic growth and exponential growth followed by decay. The former is a reasonable approximation for certain chronic infections or opportunistic pathogens in immunocompromised hosts where the pathogen is not cleared (e.g., *P. aeruginosa* infections in cystic fibrosis patients), whereas the latter is a good approximation for many acute infections that are cleared due to the host's immune response (e.g., influenza, measles), or for chronic infections where the pathogen load peaks before falling to lower levels (e.g., HIV, syphilis). In both cases, we show that the transient within-host dynamics lead to substantial shifts in epidemic dynamics at the population-level compared to steady-state approximations. In the case of logistic growth, the steady-state approximation for within-host pathogen load always overestimates the transmission rate, which leads to a much faster growth rate and higher endemic prevalence. With exponential growth and decay in pathogen load, the steady-state approximation can lead to a slower growth rate and lower endemic prevalence. The latter scenario especially highlights issues with using a steady-state approximation, as the steady state for acute infections is a pathogen load of zero, which would clearly prevent any transmission. To facilitate comparisons with the steady state approximation we assumed a non-zero pathogen load, but this is not a limitation of our model (one could set the steady state pathogen load to zero but maintain the pathogen in the population due to the transient dynamics). Our framework therefore allows for acute infections with full pathogen clearance whereas the steady-state approximation does not. A steady state approximation also only allows for unidirectional feedback from within-host to between-host processes (termed "inessential" couplings in [12]) and does not allow for a full ("essential") feedback loop with between-host processes impacting on the within-host dynamics. However, the full feedback loop can readily be captured using our framework, for example by setting the initial inoculum to be a function of the transmission rate. Very few studies incorporate this full feedback loop [21,30,31], likely due to the predominance of the steady-state approximation, but we expect that our framework will offer new opportunities for exploring the effects of the full within- and between-host feedback loop on epidemic dynamics and pathogen evolution.

We further illustrated how our method can be applied to study pathogen evolution, both phenotypically (e.g., evolution of virulence) and genetically (e.g., for simulating true phylogenetic trees). As with the fundamental shifts observed in epidemiological dynamics, we showed how a steady-state approximation for within-host pathogen dynamics can lead to a random walk through trait space, but there is directional selection towards an evolutionarily stable level of virulence when the transient dynamics are fully captured. This occurs because selection acts on the growth rate of the pathogen, rather than the steady state, which is fixed. Previous theory using IDEs in a few specific scenarios has shown that virulence evolution can be greatly affected by within-host dynamics, which are difficult to capture using conventional methods. These include the order and timing of co-infecting pathogens [30], contrasting selection pressures at the within- and between-host levels [24] and the strength of the host immune response [32]. As our modelling framework does not rely on IDEs, which greatly limit the complexity of within-host dynamics, our method will significantly expand the scope for studying pathogen evolution across scales.

In addition to the phenotypic evolution of key disease traits such as virulence, we have also shown how our framework can be used to efficiently simulate the molecular evolution of pathogens. A major strength of our approach is that infections and transmission dynamics are tracked at the individual level, allowing for the construction of true phylogenetic and transmission trees. This not only offers a novel method to aid in the development of tree reconstruction methods, but would allow, for example, one to explore how different processes at the within-host level impact on phylogenetic patterns that are observed at the population level, or vice versa. While several methods currently exist for simulating phylogenetic and transmission trees using standard epidemiological models [28,33], a unique advantage of our approach is that the modular nature of our framework allows one to readily adapt the within-host, population-level, and coupling functions to any feasible scenario. For example, the within-host dynamics could be adapted to account for coinfections of multiple pathogens, pharmacodynamics and antimicrobial resistance, interactions with the microbiome, or innate and adaptive responses, and the population-level dynamics could be modified to include migration, vaccination, or non-pharmaceutical interventions. Our framework is therefore likely to provide valuable insights across a wide range of scenarios. Furthermore, while we have focused on host-pathogen interactions our method could readily be applied to model other constituents of the host microbiome, including horizontally or vertically transmitted symbionts [13,34] or hyperparasites [35].

Capturing the full complexity of real populations, including all processes at the within-host level, is neither attainable nor desirable. Yet capturing key aspects of within-host pathogen and immune dynamics is crucial for a better understanding of pathogen epidemiology and evolution. While steady-state approximations offer some insights into the impact of within-host processes on population-level dynamics, fully capturing transient interactions between pathogens, host immune responses, and the microbiome is a major challenge that will shed light on pathogen ecology and evolution across scales. Our framework provides a novel, readily adaptable approach for coupling a wide range of within-host and population-level dynamics that is both efficient and realistic.

# Efficient coupling of within- and between-host infectious disease dynamics

Cameron A. Smith[1,4,*] and Ben Ashby[2,3,4]

1. Department of Biology, University of Oxford, Oxford, UK
2. Department of Mathematics, Simon Fraser University, Burnaby, BC, Canada
3. Pacific Institute on Pathogens, Pandemics and Society, Simon Fraser University, Burnaby, BC, Canada
4. Department of Mathematics, University of Bath
*. Corresponding author: cameron.smith@biology.ox.ac.uk

**Abstract**

Mathematical models of infectious disease transmission typically neglect within-host dynamics. Yet within-host dynamics – including pathogen replication, host immune responses, and interactions with microbiota – are crucial not only for determining the progression of disease at the individual level, but also for driving within-host evolution and onwards transmission, and therefore shape dynamics at the population level. Various approaches have been proposed to model both within- and between-host dynamics, but these typically require considerable simplifying assumptions to couple processes at contrasting scales (e.g., the within-host dynamics quickly reach a steady state) or are computationally intensive. Here we propose a novel, readily adaptable and broadly applicable method for modelling both within- and between-host processes which can fully couple dynamics across scales and is both accurate and computationally efficient. By individually tracking the deterministic within-host dynamics of infected individuals, and stochastically coupling these to continuous host state variables at the population-level, we take advantage of fast numerical methods at both scales while still capturing individual transient within-host dynamics and stochasticity in transmission between hosts. Our approach closely agrees with full stochastic individual-based simulations and is especially useful when the within-host dynamics do not rapidly reach a steady state or over longer timescales to track pathogen evolution. By applying our method to different pathogen growth scenarios we show how common simplifying assumptions fundamentally change epidemiological and evolutionary dynamics.

# S1 Full hybrid algorithm

The full hybrid algorithm is shown below. All notation is the same as in the full text. Note that steps (1c), (1d) and (1e) can be done in any order, or randomly at every time step.

**Algorithm 1: Hybrid method**

**INPUT:** Within-host dynamics — $\boldsymbol{f}_W(\boldsymbol{W}, \boldsymbol{\theta}_W, T)$; Host-demographics dynamics — $\boldsymbol{f}_D(\boldsymbol{D}, \boldsymbol{\theta}_D, t)$; Transmission coupling function — $\beta(\boldsymbol{W}, \boldsymbol{\theta}_C, T)$; Mortality coupling function — $\alpha(\boldsymbol{W}, \boldsymbol{\theta}_C, T)$; Recovery coupling function $\gamma(\boldsymbol{W}, \boldsymbol{\theta}_C, T)$; Initial number of infected individuals — $N_I(0)$; Initial within-host states — $\boldsymbol{W}_i(\boldsymbol{\theta}_W, 0)$ for $i \in \{1, ..., N_I(0)\}$; Initial population-level state — $[S(0), I(0), R(0)]$; Time-step for simulation — $\delta t$; Normalisation constant — $A$

At time $t > 0$:

**(1a)** Update the host demographics by updating the solution over $[t, t + \delta t]$.

**(1b)** If necessary, probabilistically remove an infected individual to ensure that, on average, the number of individuals matches the density.

1. Draw a uniform random number $u \sim Uniform(0, 1)$.
2. If $u < N_I(t) - AI(t)$, remove an individual uniformly at random. This is treated as a background mortality.
3. Update $N_I(t) \leftarrow N_I(t) - 1$.

**(1c)** For each infected individual $i \in \{1, ..., N_I(t)\}$, enact any transmission events.

1. Calculate the number of individuals that the individual infects in $[t, t + \delta t)$ by sampling $n_i \sim Poisson(\beta(\boldsymbol{W}_i, \boldsymbol{\theta}_C, T_i)\delta t)$.
2. Generate $n_i$ new infected individuals with initial within-host state $\boldsymbol{W}_j(\boldsymbol{\theta}_W, 0)$, where $j$ indexes the new infected individuals.
3. Update the population-level state variables by updating $S(t+\delta t) \leftarrow S(t+\delta t) - n_i/A$ and $I(t+\delta t) \leftarrow I(t+\delta t) + n_i/A$.
4. Update $N_I(t) \leftarrow N_I(t) + n_i$.

**(1d)** For each infected individual $j \in \{1, ..., N_I(t)\}$, enact a possible recovery event, if applicable.

1. Calculate the probability of recovery in $[t, t + \delta t)$ by drawing $u_{r,j} \sim Uniform(0, 1)$. If $u_{r,j} < \gamma(\boldsymbol{W}_j, \boldsymbol{\theta}_C, T_j)\delta t$, then a recovery event occurs.
2. If a recovery event occurs, update the population-level state variables $I(t + \delta t) \leftarrow I(t + \delta t) - 1/A$ and $R(t + \delta t) \leftarrow R(t + \delta t) + 1/A$. Update the number of individuals $N_I(t) \leftarrow N_I(t) - 1$.

**(1e)** For each infected individual $k \in \{1, ..., N_I(t)\}$, enact a possible mortality event.

1. Calculate the probability of mortality in $[t, t + \delta t)$ by drawing $u_{v,k} \sim Uniform(0, 1)$. If $u_{v,k} < \alpha(\boldsymbol{W}_k, \boldsymbol{\theta}_C, T_k)\delta t$, then a mortality occurs.
2. If a mortality occurs, update the population-level state variables $I(t+\delta t) \leftarrow I(t+\delta t) - 1/A$. Update the number of individuals $N_I(t) \leftarrow N_I(t) - 1$.

**(1f)** Update the within-host dynamics for all infected individuals over $[t, t + \delta t)$. Update the infection ages as $T_i \leftarrow T_i + \delta t$ for $i \in \{1, ..., N_I(t)\}$.

# S2 Example 1: Logistic within-host dynamics

In this section, we review the first example, which is used to model infections that are not cleared by the immune system.

## S2.1 Host-demographic dynamics (HDD)

We begin by describing the HDD. We define the density of susceptible individuals to be $S(t)$ and the density of infected individuals to be $I(t)$. We will assume that once infected, individuals do not recover. Let $b$ be the per-capita host growth rate (units of per day) and $q$ (units of per day per individual) denote density dependent effects on host growth. Further, define $d$ (units of per day) to be the density-dependent background mortality, and $A$ the scaling constant from the main text, which allows scales the carrying capacity. Then our HDDs are:

$$\frac{dS}{dt} = N(b - AqN) - dS, \tag{S1}$$

$$\frac{dI}{dt} = -dI, \tag{S2}$$

$$N = S + I. \tag{S3}$$

## S2.2 Within-host dynamics (WHD)

We now define the WHD. We will assume that the pathogen load in all individuals grows logistically to some capacity $K$ (units of pathogen count), and with growth rate $r$ (units of per day). If we define $T_i$ to be the infection age of an individual $i$, then $P_i(T_i)$ is the pathogen load of individual $i$ at infection age $T_i$, with:

$$\frac{dP_i}{dT_i} = rP_i\left(1 - \frac{P_i}{K}\right), \qquad P(0) = P_0. \tag{S4}$$

## S2.3 Coupling functions

Finally, we define the coupling functions. Since there is no recovery in this example, we only need to define transmission and virulence. For simplicity, we assume that transmission is proportional to pathogen load, but virulence accelerates as the square of the pathogen load. If we let $\hat{\beta}$ and $\hat{\alpha}$ be scaling parameters for the transmission and virulence respectively, then:

$$\beta(P_i) = \hat{\beta} P_i, \tag{S5}$$

$$\alpha(P_i) = \hat{\alpha} P_i^2. \tag{S6}$$

## S2.4 Evolution

For Fig. 4 of the main text, we conduct an evolutionary simulation to demonstrate that our novel hybrid method can be applied to investigate evolution. We do this using the logistic example from Section S2, which requires a change in notation. The evolving trait will be the pathogen growth rate, $r_j$.

We choose a set of pathogen replication rates $\{r_j : j = 0, ..., n_{r-1}\}$, where $r_0 = 0$ and $r_j = jh$, where $h$ is the mutation effect size. Then we write $P_{ij}(T_i)$ to be the pathogen load for individual $i$ and pathogen growth rate $r_j$, and which satisfies the ODE:

$$\frac{dP_{ij}}{dT_i} = r_j P_i\left(1 - \frac{P_i}{K}\right). \tag{S7}$$

For the HDD, instead of having a single infected class, there are $n_r$ infected classes $I_j$ for $j \in \{0, ..., n_{r-1}\}$ which describe the density of hosts that are infected with pathogen strain $r_j$. We then have a system of HDD equations:

$$\frac{dS}{dt} = N(b - AqN) - dS,$$

$$\frac{dI_j}{dt} = -dI_j,$$

$$N = S + \sum_{j=0}^{n_r-1} I_j.$$

The algorithm proceeds as follows. We begin by initialising a single pathogen strain $j$, and running the hybrid algorithm for 100 time units. We then introduce a mutation by choosing a neighbouring strain (in this case $j-1$ or $j+1$) with equal probability. Note that if more than one strain is present in the population, we firstly need to choose which of these mutates, which we do with probability proportional to the prevalence of each strain. Once a mutant strain has been chosen, we take one of our infected individuals with the original strain ($r_j$) and update it to be the mutant. We then repeat this process over multiple evolutionary time steps.

| Parameter | Value | Description | Note |
|---|---|---|---|
| $b$ | 0.5 | Birth rate | - |
| $q$ | $2.475 \times 10^{-3}$ | Density-dependent strength | Chosen so that the total number of individuals is 200 at disease-free steady state |
| $d$ | $5 \times 10^{-3}$ | Background mortality | - |
| $r$ | 2.5 | Pathogen growth rate | - |
| $K$ | $3 \times 10^9$ | Pathogen carrying capacity | - |
| $\hat{\beta}$ | $10^{-10}$ | Transmission parameter | - |
| $\hat{\alpha}$ | $5 \times 10^{-21}$ | Virulence parameter | - |

Table S1: Parameters used in simulations for Figs. 2 and 4 of the main text.

## S3 Example 2: Pathogen load growth and decay

For this section, we introduce the various scales for the second example, where pathogen load peaks and then decays. This is a reasonable approximation for many acute infections with recovery.

### S3.1 Host-demographic dynamics (HDDs)

We define $S(t)$ and $I(t)$ as above, and define $R(t)$ to be the density of recovered individual at time $t$. Using the same notation from Section S2.1 we obtain the following ODEs for the host demographics:

$$\frac{dS}{dt} = N(b - AqN) - dS, \tag{S8}$$

$$\frac{dI}{dt} = -dI, \tag{S9}$$

$$\frac{dR}{dt} = -dR, \tag{S10}$$

$$N = S + I + R. \tag{S11}$$

### S3.2 Within-host dynamics (WHDs)

At the within-host level, we require a functional form which peaks and then decays. We use the following ODE for the pathogen load for individual $i$ at infection age $T_i$, given by $P_i(T_i)$:

$$\frac{dP_i}{dT_i} = re^{-\eta T_i} + \eta\left(P^* - P_i\right), \qquad P_i(0) = P_0. \tag{S12}$$

Here, $P^*$ is the steady state of the system (which may be zero, or a low-level background pathogen load), and $r, \eta$ are positive parameters that control the growth and decay rates. Note that early in an infection, the right-hand side of the ODE is positive and so the pathogen grows in number. As the pathogen grows in number, the $-\eta P_i$ term will begin to dominate, causing the growth in pathogen load to slow and then decay to $P^*$.

This system yields an explicit function which describes the pathogen load at time $T_i$:

$$P_i(T_i) = r\left(T_i - \frac{P^* - P_0}{r}\right)e^{-\eta T_i} + P^*. \tag{S13}$$

From this solution, we can calculate that the peak pathogen load occurs at time

$$T_M = \frac{1}{\eta} + \frac{P^* - P_0}{r}, \tag{S14}$$

and takes the value

$$P_M := P(T_M) = \frac{r}{\eta}\exp\left\{-\left(1 + \frac{\eta}{r}(P^* - P_0)\right)\right\}. \tag{S15}$$

### S3.3 Coupling functions

Finally, we define the coupling functions. For both transmission and virulence, we use the same functional forms as in Section S2.3. For recovery, we assume that the rate of recovery decreases exponentially as the pathogen load increases. The decay in pathogen load arises from an implicit immune response. We specify three parameters, $P_T$ which is a threshold pathogen count, $\hat{\gamma}_B$ which is the background rate of recovery when the pathogen count reaches the threshold, and $\hat{\gamma}_S$ which is a scaling parameter. Therefore we have he following coupling functions:

$$\beta(P_i) = \hat{\beta}P_i,$$
$$\alpha(P_i) = \hat{\alpha}P_i^2,$$
$$\gamma(P_i) = \hat{\gamma}_B \exp\{-\hat{\gamma}_S(P_i - P_T)\}.$$

### S3.4 Phylogenetics

In order to plot the phylogenetic tree in Fig. 5, we need to track which pathogen strain infects each host. Every strain mutates according to the following rules. In each time-step of size $\Delta$), there is a probability of $\mu\Delta$ that a locus mutates. A locus is chosen probabilistically according to the distribution $\pi_s(m)$ where $m$ is the site index, and then mutates according to a probability distribution $\pi_\ell(k|k_b)$, where $k$ is the label for the new locus value and $k_b$ is the previous label. In the simplest possible case:

$$\pi_s(m) = 1/n_s, \qquad m \in \{1, ..., n_s\}, \tag{S16}$$
$$\pi_\ell(k|k_b) = 1/3, \qquad k_b \in \{A, C, G, T\}, k \in \{A, C, G, T\}\setminus\{k_b\}. \tag{S17}$$

The first of these says that each locus is chosen uniformly at random, while the second equation states that the current nucleotide can mutate into any of the other three uniformly at random. Once we have the strain, we are able to record times of mutation and reconstruct a phylogenetic tree as seen in Fig. 5 of the main text.

## S4 Efficiency

We compare our hybrid method with a method by which every component — the WHD, HDD and coupling — are simulated using tau-leaping (Gillespie, 2001). The following simulations were conducted and timed on a Dell XPS 13 with an 11th Gen Intel Core i7 processor, and 16GB of RAM. The code was executed using Python 3 through Visual Studio Code version 1.8.2. The times per repeat are displayed in Table S4, and are plotted in Fig. S4. We see that in all combinations of population size and WHD model, the hybrid method is faster than the fully tau-leaping simulation, with the biggest savings seen in the logistic example.

| Parameter | Value | Description | Note |
|---|---|---|---|
| $b$ | 0.5 | Birth rate | - |
| $q$ | $2.475 \times 10^{-3}$ | Density-dependent strength | Chosen so that the total number of individuals is 200 at disease-free steady state |
| $d$ | $5 \times 10^{-3}$ | Background mortality | - |
| $r$ | $10^9$ | Pathogen scaling parameter | - |
| $\eta$ | $1/3$ | Exponential decay rate | - |
| $P^*$ | 100 | Pathogen steady state | - |
| $\hat{\beta}$ | $10^{-10}$ | Transmission parameter | - |
| $\hat{\alpha}$ | $5 \times 10^{-21}$ | Virulence parameter | - |
| $\hat{\gamma}_1$ | 0.1 | Recovery parameter | - |
| $\hat{\gamma}_2$ | $10^{-6}$ | Recovery parameter | - |

Table S2: Parameters used in simulations for Figs. 3 and 5 of the main text.

| $N$ | WHD model | Simulation type | Time per repeat (seconds) | Percentage of tau-leaping (%) |
|---|---|---|---|---|
| 100 | Logistic | Hybrid | 0.3115 | 33.25 |
| | | Tau-leaping | 0.9368 | — |
| | Peaked | Hybrid | 0.3794 | 58.13 |
| | | Tau-leaping | 0.6527 | — |
| 200 | Logistic | Hybrid | 0.5404 | 29.68 |
| | | Tau-leaping | 1.8208 | — |
| | Peaked | Hybrid | 0.8868 | 63.70 |
| | | Tau-leaping | 1.3921 | — |
| 500 | Logistic | Hybrid | 1.3175 | 28.24 |
| | | Tau-leaping | 4.6653 | — |
| | Peaked | Hybrid | 2.2714 | 59.47 |
| | | Tau-leaping | 3.8192 | — |
| 1000 | Logistic | Hybrid | 3.0312 | 28.39 |
| | | Tau-leaping | 10.6772 | — |
| | Peaked | Hybrid | 5.3724 | 53.27 |
| | | Tau-leaping | 10.0844 | — |

Table S3: **Table of timings for four different combinations of model and simulation type.** We average over 500 repeats and simulate for 4 different values of population size. The two models are the examples from the main text, while the simulation types are hybrid or full tau-leaping, which uses tau-leaping at all three scales.

# References

D.T. Gillespie. Approximate accelerated stochastic simulation of chemically reacting systems. *J. Chem. Phys.*, 115(4):1716–1733, 2001.

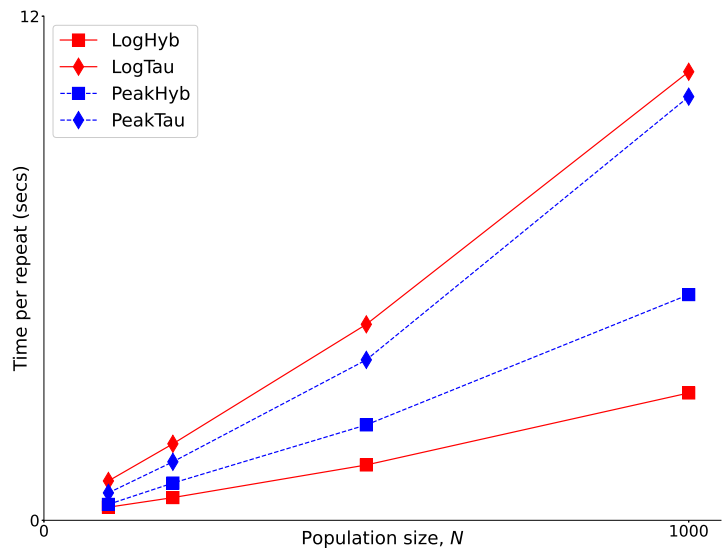

Figure S1: **Efficiency for the hybrid method.** The efficiency of the two two examples from the main text, under two simulation techniques. Red solid lines are the logistic example, blue dashed lines are the peaked example. Square markers denote the hybrid method, while the diamond markers are a full tau-leaping method.